\documentclass[unnumsec,webpdf,contemporary,large]{oup-authoring-template}%

\graphicspath{{Fig/}}

\usepackage{hyperref}

\theoremstyle{thmstyleone}%
\theoremstyle{thmstyletwo}%
\theoremstyle{thmstylethree}%

\begin{document}

\journaltitle{Journal}
\DOI{DOI added during production}
\copyrightyear{YEAR}
\pubyear{YEAR}
\vol{XX}
\issue{x}
\access{Published: Date added during production}
\appnotes{Application Note | Systems Biology}

\firstpage{1}


\title[PEtab SciML]{PEtab SciML: an exchange format for specifying and training dynamic scientific machine learning models}

\author[1]{Sebastian Persson \ORCID{0009-0001-2304-4263}}
\author[1]{Branwen Snelling \ORCID{0000-0002-7205-5516}}
\author[2,3]{Maren Philipps \ORCID{0000-0001-5456-6104}}
\author[2,3]{Daniel Weindl \ORCID{0000-0001-9963-6057}}
\author[4,5]{Marija Cvijovic \ORCID{0000-0002-5142-5100}}
\author[2,3]{Jan Hasenauer \ORCID{0000-0002-4935-3312}}
\author[2,3$\dagger$]{Dilan Pathirana \ORCID{0000-0001-7000-2659}}
\author[1$\dagger$$\ast$]{Fabian Fröhlich\ORCID{0000-0002-5360-4292}}

\address[1]{\orgdiv{Dynamics of Living Systems Laboratory}, \orgname{The Francis Crick Institute}, \orgaddress{\street{1 Midland Rd}, \postcode{NW1 1AT}, \state{London}, \country{United Kingdom}}}
\address[2]{\orgdiv{Bonn Center for Mathematical Life Sciences}, \orgname{University of Bonn}, \orgaddress{\street{Regina-Pacis-Weg 3}, \postcode{53113}, \state{NRW}, \country{Germany}}}
\address[3]{\orgdiv{Life and Medical Sciences (LIMES) Institute}, \orgname{University of Bonn}, \orgaddress{\street{Regina-Pacis-Weg 3}, \postcode{53113}, \state{NRW}, \country{Germany}}}
\address[4]{\orgdiv{Department of Mathematical Sciences}, \orgname{Chalmers University of Technology}, \orgaddress{\street{Chalmersplatsen 1}, \postcode{412 96}, \state{Gothenburg}, \country{Sweden}}}
\address[5]{\orgdiv{Department of Mathematical Sciences}, \orgname{University of Gothenburg}, \orgaddress{\street{Universitetsplatsen 1}, \postcode{405 30}, \state{Gothenburg}, \country{Sweden}}}

\corresp[$\dagger$]{These authors share senior authorship }
\corresp[$\ast$]{Corresponding author. \href{fabian.frohlich@crick.ac.uk}{fabian.frohlich@crick.ac.uk}}

\received{Date}{0}{Year}
\revised{Date}{0}{Year}
\accepted{Date}{0}{Year}



\abstract{\normalfont
\textbf{Summary:}
Dynamic scientific machine learning (SciML) models that combine mechanistic ordinary differential equations (ODEs) with machine-learning (ML) components have applications ranging from learning unknown biological processes to integrating auxiliary data modalities into dynamic modelling. To enable reproducible and efficient SciML training, we introduce \textbf{PEtab SciML}, an interoperable data format for specifying parameter-estimation problems in which mechanistic and ML model parameters are jointly estimated from time-series data. PEtab SciML supports several ML--ODE hybridization patterns in realistic problem setups. It is accompanied by a reference Python library and downstream modelling support in Python/JAX and Julia, provided by \texttt{AMICI} and \texttt{PEtab.jl}, respectively, and a collection of real-data benchmarks.\newline
\textbf{Availability and implementation:} PEtab SciML is available on GitHub (\url{https://github.com/PEtab-dev/petab_sciml}). The reference Python package is installable from \texttt{PyPI} and is continuously tested and supported on Linux, macOS, and Windows.
}


\maketitle

\section{Introduction}

\begin{figure*}[t]
    \centering
    \includegraphics[width=0.95\linewidth]{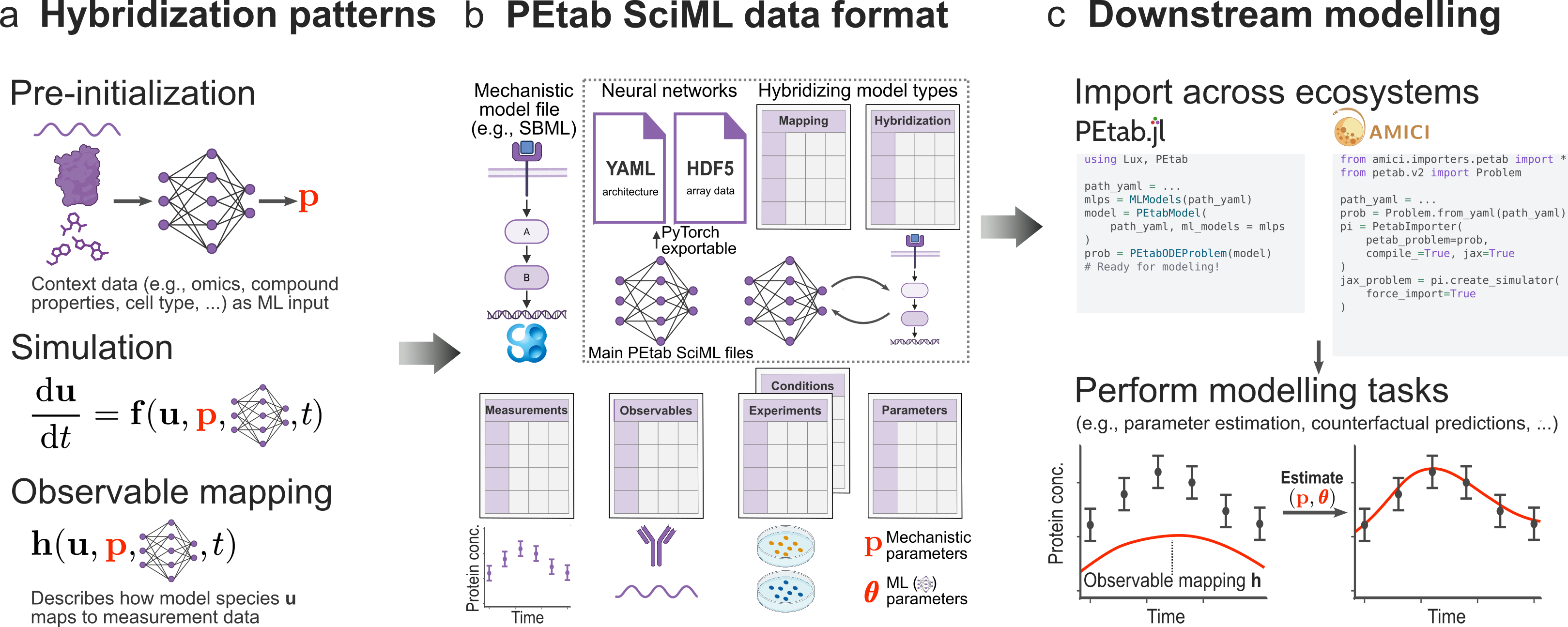}
    \caption{
    \textbf{PEtab SciML: an exchangeable data format for SciML modelling.}
    \textbf{(a)} PEtab SciML supports three main hybridization patterns for combining mechanistic and ML models, including combinations within a single problem.
    \textbf{(b)} As a PEtab v2 extension, PEtab SciML retains the core PEtab model file(s) and tables, thereby supporting all PEtab v2 problem features such as multiple observables and/or experimental conditions. It further introduces files and table extensions for specifying neural network models and their hybridization patterns.
    \textbf{(c)} PEtab SciML problems can be imported into modelling packages across computational ecosystems for downstream tasks such as simulation and training.
    Figure partially created with BioRender.com.}
    \label{fig:1}
\end{figure*}

Mechanistic ordinary differential equation (ODE) models are widely used to understand the dynamics of biological processes. However, purely mechanistic models are limited by an incomplete understanding of the underlying mechanisms and by gaps or inconsistencies in the literature. In addition, integrating informative non-time-series data modalities, such as omics data and drug compound properties, to improve predictive performance remains challenging. This has motivated the use of scientific machine learning (SciML) models that hybridize ODE and machine learning (ML) models~\citep{noordijk_rise_2024, metzcar_review_2024}. Such hybridization can take several forms, including ML models embedded in the ODE right-hand side to learn hard-to-model processes~\citep{rackauckas_universal_2021}, ML models placed upstream of the ODE to map informative data modalities to model parameters~\citep{fabrini2026, beckers_deepct_2024}, and ML models in the observation mapping to account for model misspecification~\citep{hass_predicting_2017} (Figure~\ref{fig:1}a). Reflecting their applicability, SciML models are increasingly used across disciplines such as systems biology~\citep{rooij_physiology-informed_2025}, immunology~\citep{dandekar_machine_2020, ye_integrating_2025}, and pharmacology~\citep{beckers_deepct_2024, bram_low-dimensional_2024}.

To be predictive, SciML models must typically be trained against experimental time-series data by estimating unknown model parameters. Ideally, ML and ODE model parameters are estimated jointly in an end-to-end formulation, so that all model components are learned together. However, implementing such training is challenging at both the specification and numerical levels. At the specification level, the training objective must represent the experimental design and observation process. In biology, such data are often sparse, measure only a subset of model species, and are collected across multiple experimental conditions such as different drug dosing schemes~\citep{hass_benchmark_2019}. At the numerical level, ODE models often exhibit hard-to-simulate stiff dynamics~\citep{persson_petabjl_2025}. In addition, SciML models contain many parameters through their ML modules, making scalable gradient-computation methods essential for runtime-efficient training~\citep{kidger_neural_2022}. These challenges create a need for user-friendly specification of SciML training problems and performant software.

SciML software ecosystems, such as Julia SciML~\citep{rackauckas_universal_2021} and the Python/JAX-based \texttt{equinox}~\citep{kidger_neural_2022}, provide building blocks for implementing SciML training. However, modellers must still implement non-trivial training objectives that account for both specification-level features, like multiple experimental conditions, and numerical aspects, like efficient gradient computation. Such bespoke implementations are time-consuming and error-prone, limiting both reproducibility and exchangeability.
PEtab, which is a standardized exchange format with a broad range of features for specifying parameter-estimation problems in systems biology~\citep{schmiester_petabinteroperable_2021}, helps address these challenges for mechanistic ODE models. Parameter estimation problems in the PEtab format can be reproducibly exchanged and imported into established software packages across programming languages.

Here, we introduce \textbf{PEtab SciML}, a PEtab v2~\citep{pathirana_2026} extension for specifying parameter-estimation problems for SciML models. PEtab SciML supports several ML--ODE hybridization patterns and, by extending PEtab, enables their specification across the broad range of realistic problem setups already supported by PEtab. To enable downstream use, we also provide import support through two established parameter-estimation packages: \texttt{AMICI} via its JAX backend~\citep{frohlich_amici_2021} and the Julia package \texttt{PEtab.jl}~\citep{persson_petabjl_2025}.

\section{The PEtab SciML data format}

PEtab SciML is an interoperable format for specifying parameter-estimation problems for SciML models (Figure~\ref{fig:1}). It targets problems where parameters of the mechanistic and ML components are jointly estimated from noisy time-series data. The aim is to simplify SciML modelling workflows and improve reproducibility by providing a consistent, easy-to-use format that downstream software packages can import for tasks ranging from training to sensitivity analysis (Figure~\ref{fig:1}b-c).

A PEtab SciML problem is a compact bundle of model file(s) in standard formats such as SBML~\citep{keating_sbml_2020}, human-readable TSV tables, ML model files, and array data files (Figure~\ref{fig:1}b). As an extension of the PEtab v2 format~\citep{pathirana_2026}, PEtab SciML retains the core PEtab TSV tables that specify model observables (which link simulations to model state), measurement data, parameters to estimate, and optional experimental conditions. Beyond the core PEtab components, specifying a SciML problem requires three additional elements: (i) the specification of the ML model, (ii) the hybridization of the ML model with the mechanistic model, and (iii) the specification of high-dimensional array data for ML model inputs and parameters. PEtab SciML defines these by introducing: a YAML-based exchange format for defining neural network models, which can be generated by exporting PyTorch modules; a hybridization table that, together with the mapping table, specifies how ODE and ML models interact; and an HDF5-based array format for storing high-dimensional array data, such as ML model inputs and/or parameters.

PEtab SciML supports several hybridization patterns for combining mechanistic and ML models (Figure~\ref{fig:1}a). These and other features are described in the online documentation through extensive tutorials. The documentation also provides the detailed format specification and serves as the most up-to-date reference. In the following subsections, we summarize selected key features.

\subsection{Hybridization patterns and supported problem features}

PEtab SciML supports three patterns for hybridizing ML and ODE models (Figure~\ref{fig:1}a). First, ML models can be placed upstream of the ODE model and evaluated before simulation. This enables informative non-time-series data modalities, such as images, patient metadata, or omics data, to be mapped to ODE model parameters and/or initial values. Second, ML models can be embedded directly in the ODE right-hand side. This covers both universal differential equations (UDEs)~\citep{rackauckas_universal_2021}, also referred to as grey-box models or hybrid neural ODEs, and neural ODEs (NODEs) in which the full ODE right-hand side is represented by a neural network~\citep{chen_neural_2018}. Third, ML models can be used in the observation mapping linking simulated outputs to measurements, for example to account for model misspecification~\citep{philipps_current_2025}. Through the PEtab SciML neural-network YAML format, these hybridizations are supported for a wide range of architectures, including feed-forward and convolutional neural networks. A single PEtab SciML problem can involve multiple hybridizations.

Because PEtab SciML extends PEtab v2, these hybridization patterns can be applied to the broad range of problem setups supported by the format. This includes problems with multiple observables and/or experimental conditions, events such as drug dosing, pre-equilibration (steady-state initialization), parameter priors for Bayesian inference, and diverse measurement noise models.

\subsection{Scope of the format}

Like PEtab~\citep{schmiester_petabinteroperable_2021}, PEtab SciML deliberately focuses on problem specification rather than prescribing optimizers, ODE solvers, or gradient methods. Specified problems are instead intended to be imported into existing modelling ecosystems through suitable packages (Figure~\ref{fig:1}c; section below). PEtab SciML also does not aim to cover all SciML problem classes. It targets problems with an explicit ODE model structure that enters the parameter-estimation problem as a hard constraint. In particular, it does not cover physics-informed neural networks (PINNs)~\citep{raissi_physics-informed_2019}, where mechanistic knowledge is typically imposed through soft constraints in the training objective, and for which dedicated software ecosystems already exist~\citep{lu_deepxde_2021}.

\section{Software ecosystem}

PEtab SciML builds on the PEtab software ecosystem, comprising a mature standard and several packages that import the format~\citep{schmiester_petabinteroperable_2021}. It extends this foundation with a Python reference library for PEtab SciML problems and dedicated importers that enable downstream workflows such as simulation and training. Currently, PEtab SciML problems can be imported through \texttt{AMICI} for Python/JAX~\citep{frohlich_amici_2021} and \texttt{PEtab.jl} for Julia~\citep{persson_petabjl_2025} (Figure~\ref{fig:1}c). Below, we briefly outline each component.

The \texttt{petab-sciml} Python library provides tools for creating, validating, and transforming SciML problems. These include linting for problem correctness, utilities for array and ML model files, helpers for creating common problem types such as NODEs, and transformations for using training strategies like multiple shooting.

\texttt{AMICI} provides a Python/JAX-based interface for importing PEtab SciML problems. During import, it generates optimized, differentiable modules built on the packages \texttt{equinox} and \texttt{diffrax}~\citep{kidger_neural_2022}. These modules enable gradients of both mechanistic and ML parameters to be computed using scalable automatic differentiation when solving the parameter-estimation problem.

\texttt{PEtab.jl} can import PEtab SciML problems in the standard format and provides a user-friendly Julia API for creating such problems, internally represented in the format. Built on the Julia SciML ecosystem, \texttt{PEtab.jl} exposes imported problems to performant ODE solvers in \texttt{DifferentialEquations.jl}~\citep{rackauckas_differentialequationsjl_2017}. It also supports multiple gradient computation methods, including forward-mode automatic differentiation and adjoint sensitivities. Once imported, problems can be used with Julia's optimization and Bayesian inference tooling.

\subsection{Test suite and benchmark collection}

A public test suite is continuously used to assess importer correctness and serves as a reference for new importers. The test suite consists of small scale PEtab SciML problems that cover the full breadth of supported features including the three different hybridization patterns, the wide range of neural network architectures, and PEtab v2 features (experimental conditions, events, pre-equilibration etc.).

Real-data benchmark collections are important for evaluating computational methodology, as they provide shared problem definitions for comparing methods under realistic conditions \citep{hass_benchmark_2019, weber_essential_2019}. To support reproducible methodological research on SciML models in areas such as training, sensitivity analysis, and identifiability analysis, we provide a benchmark collection of real-data PEtab SciML problems. The problems in the collection exhibit different hybridization patterns and applications, ranging from intracellular modelling to epidemiology. By separating problem specifications from backend-specific implementation code, the PEtab SciML format enables contributed benchmarks to be reused across computational ecosystems. Similar to the PEtab benchmark collection for ODE models~\citep{schmiester_petabinteroperable_2021}, the collection is intended to grow as new models become available.

\subsection{Enabling efficient training strategies}

Training SciML models is often challenging~\citep{philipps_current_2025}. This has motivated the use of training strategies that aim to improve optimization by reformulating the training problem. Three such strategies are curriculum learning (including approaches sometimes termed growing fits)~\citep{bengio_curriculum_2009, rackauckas_diffeqfluxjl_2019}, in which the model is trained on progressively longer time intervals; multiple shooting~\citep{turan_multiple_2022}, in which the simulation interval is divided into shorter windows with separately initialized trajectories that are coupled through continuity conditions; and curriculum multiple shooting~\citep{persson2026curriculummultipleshootingrobust}, which combines ideas from both approaches.

The PEtab SciML Python library supports these three strategies by transforming a PEtab SciML problem into one or more PEtab problems encoding the selected strategy. The transformed problem(s) can then be imported by downstream tools like any other PEtab problem. Thus, PEtab SciML not only supports the specification of SciML problems, but also represents them in a format that enables the application of efficient training strategies.

\section{Discussion}

PEtab SciML provides a programming-language-independent data format for specifying parameter-estimation problems for SciML models that combine ML with mechanistic ODE components through several hybridization patterns. Together with its Python reference library, importer support in Python/JAX and Julia, and benchmark collection, PEtab SciML simplifies the specification, sharing, and use of SciML models.

Future work will focus on extending PEtab SciML alongside the broader PEtab ecosystem. This includes incorporating features expected in future PEtab releases, such as standardized result formats for model training, and supporting additional model classes through planned PEtab extensions, like nonlinear mixed-effects models~\citep{davidian_nonlinear_2003}. Further development of PEtab SciML importers will aim to improve training speed by expanding support for emerging solver and automatic-differentiation backends, including GPU-accelerated training~\citep{utkarsh_automated_2024}. Finally, we aim to grow the collection of benchmark models to support method development for SciML modelling.

\section{Conflicts of interest}

The authors declare that they have no competing interests.

\section{Funding}

J.H. acknowledges support by the Deutsche Forschungsgemeinschaft (DFG, German Research Foundation) under Germany’s Excellence Strategy (EXC 2047 – 390685813, EXC 2151 – 390873048), the European Union through ERC grant INTEGRATE (Grant Agreement No. 101126146), and the University of Bonn (through the Schlegel Professorship). M.C acknowledges support from the Swedish  Research Council (VR2023-04319). FF, SP and BS were supported by the Francis Crick Institute, which receives its core funding from Cancer Research UK (CC2242), the UK Medical Research Council (CC2242), and the Wellcome Trust (CC2242), as well as the European Union (ERC, DeepMechanism, grant no 101163005).

The funders had no role in the design of the study; in the collection, analysis, or interpretation of data; in the writing of the manuscript; or in the decision to publish the results.

\section{Data availability}

The PEtab SciML documentation and Python reference library are available at \url{https://github.com/PEtab-dev/petab_sciml}. The PEtab SciML test suite and benchmark collection are available at \url{https://github.com/PEtab-dev/petab_sciml_testsuite} and \url{https://github.com/sebapersson/Benchmark-Models-PEtab-SciML} respectively. \texttt{AMICI} and \texttt{PEtab.jl} are open-source packages, and are available as described in their corresponding publications~\citep{frohlich_amici_2021, persson_petabjl_2025}.

\section{Author contributions statement}

Conceptualization: S.P., D.P., F.F. Methodology: S.P., B.S., M.P., D.P., F.F. Software: S.P., B.S., M.P., D.W., D.P., F.F. Writing: all authors. Supervision: M.C., J.H., D.P., F.F.

\bibliographystyle{oup-abbrvnat-scimed-authoryear}
\bibliography{PEtabSciML}



\end{document}